# ANALYSIS OF THE CHAOS DYNAMICS IN ($X_n$, $X_{n+1}$) PLANE

**Soegianto Soelistiono[1], The Houw Liong[2]**

**Bandung Institute of Technology (ITB)**
**Indonesia**
*soegianto@student.fi.itb.ac.id*

*Abstract*

In the last decade, studies of chaotic system are more often used for classical chaotic system than for quantum chaotic system. There are many ways of observing the chaotic system such as analyzing the frequency with Fourier transform or analyze initial condition distance with Liapunov exponent. This paper explains dynamic chaotic process by observing trajectory of dynamic system in ($x_n$, $x_{n+1}$) plan.

---

[1] Lab.Fisika Theory dan Komputasi, Fisika UNAIR, Physics Department, ITB, Indonesia
[2] Lecturer and Professor of Physics, Physics Department, ITB, Indonesia

## INTRODUCTION

This paper tries to explain some results of researches in answering the question of " what is chaos? ". Chaos as characteristic of a system is defined as "the system that is susceptible to initial condition so that the result is too much unpredictable".

This research tries to define the cause of chaotic system. To answer the question, the investigators consider (Xn, $X_{n+1}$) plane.

## ($X_n$, $X_{n+1}$) PLANE IS TOOL TO INVESTIGATE THE ITERATION OF SYSTEM DYNAMICS

For explanation convenience, the dynamic graph from that iterating process is used. Consider the iteration process for the logistic mapping:

$$X_{n+1} = AX_n(1 - X_n) \quad (1)$$

for $X_1$ = 0,2 and A = 3

After 20 times iteration the values are obtained:

| n | Value |
|---|---|
| 1 | 0,2 |
| 2 | 0,48 |
| 3 | 0,7488 |
| 4 | 0,56429568 |
| 5 | 0,737598197 |
| 6 | 0,580641291 |
| 7 | 0,730490947 |
| 8 | 0,590621771 |
| 9 | 0,725363084 |
| 10 | 0,597634441 |
| 11 | 0,721402548 |
| 12 | 0,602942736 |
| 13 | 0,71820838 |
| 14 | 0,607155309 |
| 15 | 0,715553219 |
| 16 | 0,610610429 |
| 17 | 0,713295999 |
| 18 | 0,613514451 |
| 19 | 0,711343409 |
| 20 | 0,616001891 |

With that above number, it is difficult to understand behavior that system. Explanation with ($x_n$, $x_{n+1}$) plane is to help understanding the behaviors of system. (See picture 1).

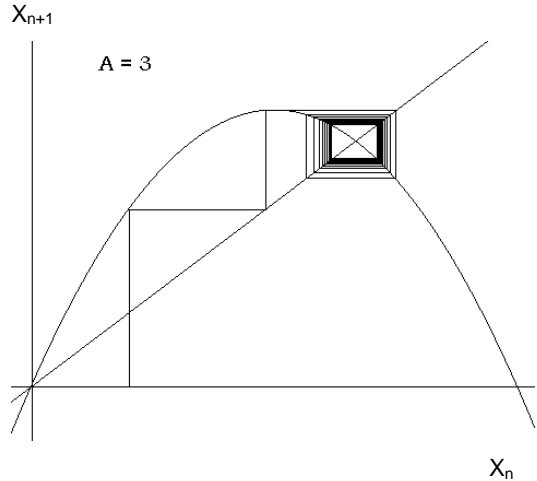

**Picture 1**. visualization of dynamic iteration in ($X_n$, $X_{n+1}$) plane

To clarify the connecting line is drawn between dot and add line $X_n = X_{n+1}$ and line of the function

F(x) = 3 x ( 1- x)

With that picture the dynamical of the iteration is known.

After this the curve F(x) is called *iterated function curve* and line $X_n = X_{n+1\,c}$ is called *reversal line* .

## GUIDANCE DYNAMICAL SYSTEM IN ($X_n$, $X_{n+1}$) PLANE

The research tries to understand the behavior of chaotic system in ( $x_{n+1}$ , $x_n$ ) plane and discussion in this paper is on theoretical basis not experimental for ODE and PDE that is well-known having chaotic system

The result will estimate that:

- The system dynamics will be guided to move toward the point or the region.
  After the system dynamics is moved into this point or region then the system dynamics cannot move to another place.

- The point or the regions that attract the dynamic system are called *attractor*.
- The System will move dynamically and continuously since there's no point or region to attract.
  In this case, the system dynamics is called the *chaotic system*.
- The system dynamics will be guided to repel and escape to infinity.
  This condition is not good because no information can be obtained.
- The system dynamics will be guided to move in circular motion like planetary motion, gyrating electron, etc.

In reality, details of system dynamics guidance are:

1. The system dynamics will be guided to move down if the iterating function curve is located under reversal line

   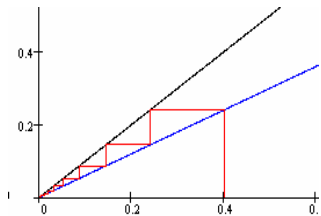

2. The system dynamics will be guided to move up if iterating function curve is located upper reversal line.

   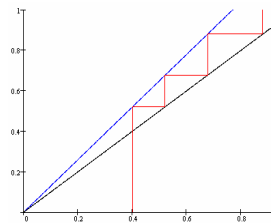

3. The dynamic system will be guide to toward crossing point if iterated function curve cross the reversal line, and satisfy

   $$0 > \frac{x_{n+2} - x_{n+1}}{x_{n+1} - x_n} > -1$$

   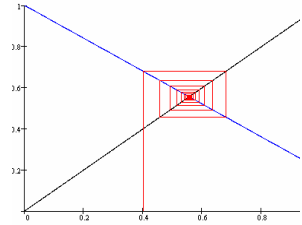

4. The dynamic system will be guided to repel the crossing point if the iterating function curve is crossing the reversal line, and satisfy $\frac{x_{n+2} - x_{n+1}}{x_{n+1} - x_n} < -1$

   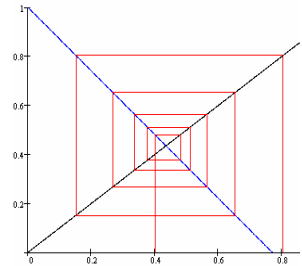

That rule of guidance system dynamics will be used to analysis the logistic equation (equation 1).

- **$0 < A < 1$ Case**

  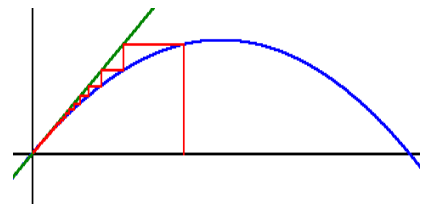

  Because iterating function curve is under the reversal line, so the system dynamics will be guided to move down toward the crossing point (attractor) that is (0,0).

- **$1 < A < 3$ Case:**

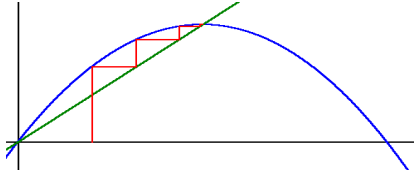

Because iterating function curve is located above the reversal line so that the system dynamics will be guided to move up toward the crossing point that is called attractor.

- **$3 < A < 3.449...$ Case:**

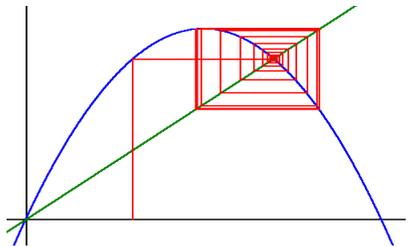

In the of case the curve likes above, there are two rule of guidance: *first* iterating function curve that is located above reversal line. The system dynamics will be guided to move up and second iterating function curve that is crossing reversal line and satisfy

$$0 > \frac{x_{n+2} - x_{n+1}}{x_{n+1} - x_n} > -1$$

That system dynamics will be guided to move forward to crossing point.

That appeared solution is **1 - 1/A**

- **$3.449... < A < 3.5699...$ Case:**

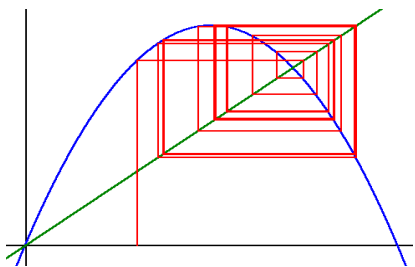

In the case of above curve, there are also two rule of guidance: *first* iterating function curve that is located above reversal line so that system dynamics will be guided to move up and *second* iterating function curve that is crossing reversal line that and

$$\frac{x_{n+2} - x_{n+1}}{x_{n+1} - x_n} < -1$$

That system dynamics will be guided to repel crossing point.

- **$3.5699... < A < 4$ Case:**

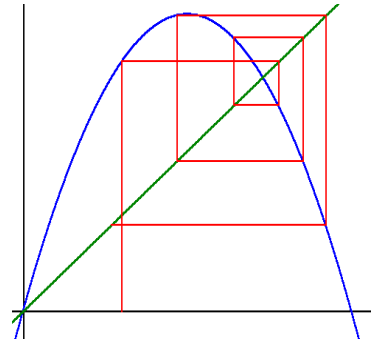

In the case of above curve, there's no attractor to attract the system dynamics so that system dynamics will keep on moving

Using guidance rule, system dynamics can define attractor and chaos, that is:

**Attractor** is the point or region that attract the system dynamics, if the system dynamics is trapped into this point or region, the system dynamics will never move anywhere else.

**Chaos** is a condition that there's no attractor to attract the system dynamics, system will keep moving.

With the above definition, equation of chaotic system can be constructed that is equation with no attractor.

Example 1 :

$$f(x) = \begin{cases} (2x - 2x^2)..........for.....x < 0.4 \\ (0,5 - x)..........if..else \end{cases}$$

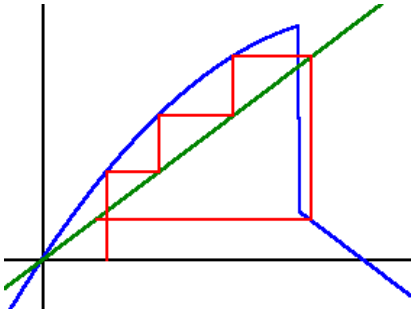

Time series the dynamical above is

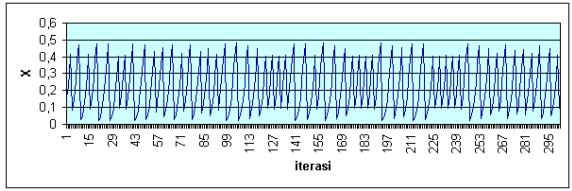

Example 2 :

$$f(x) = \begin{cases} (-0.1 + 1.2x^2) & \text{for } x > 0.3 \\ (0.5 - x) & \text{if ..else} \end{cases}$$

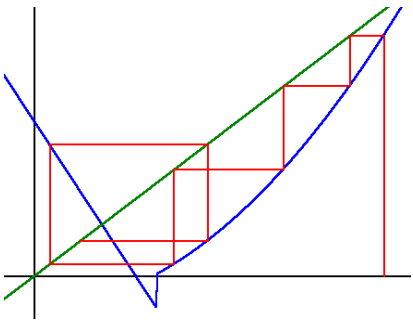

Time series the dynamical above is

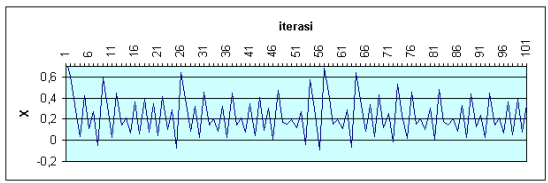

From the above explanation, it can be shown that iteration can produce chaos. Therefore, one should be careful when using iteration process.

**Solving ODE using Numerical Analysis in Chaotic System**

In nature, physical phenomena should be formulated using differential equations. It is often difficult to solve differential equation using analytical method. Another methods are using numerical computations that use iterations. As it is understood that iteration can yield chaotic solution.

One example below can be solved by using analytical method that generate exact solution. Using some numerical method for the same problem produce chaos.

Example : Consider the equation

$$\frac{dy(t)}{dt} = a.y - b.y^2 \quad (2)$$

with a and b as constant the exact solution can be obtained

$$y = \frac{a.e^{a.t}}{a + b.e^{a.t}} \quad (3)$$

The above ODE can be re-written by using **Euler** numerical solution as :

$$\frac{y_{n+1} - y_n}{h} = a.y_n - b.y_n^2 \quad (4)$$

Equation (4) also can be re-written as

$$y_{n+1} = \alpha.y_n + \beta.y_n^2 \quad (5)$$

with :

$$\alpha = a.h + 1 \quad \beta = b.h \quad (6)$$

if $\alpha = \beta = A$ then equation (5) can be written :

$$y_{n+1} = A \cdot y_n (1 - y_n) \qquad (7)$$

equation (7) is the same as equation (1), and it produces chaotic dynamics. although it is known to have exact solution (equation 3). This numerical solution is not influenced by h value, because h is only used for $\alpha$ and $\beta$.

Suppose it is chosen a = 300 and b = 100 + a, with h = 0.01,

Using the numerical Euler method

Profile (y(n),y(n+1))

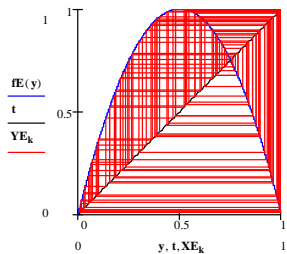

Time series

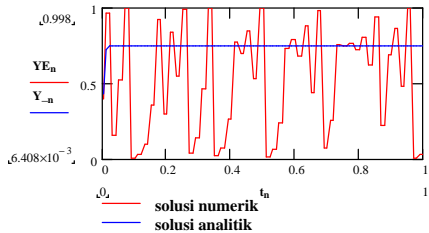

Using predictor corrector method

Profile (y(n),y(n+1))

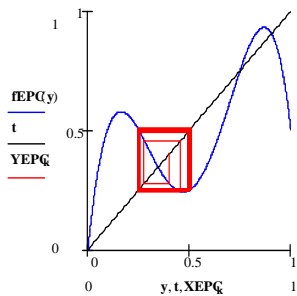

Time series

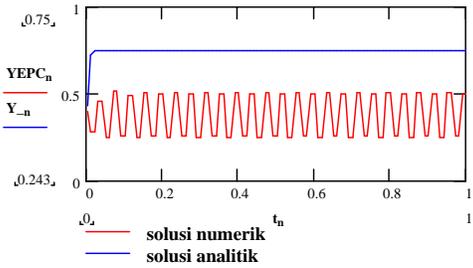

Using Runge - Kutta method

Profile (y(n),y(n+1))

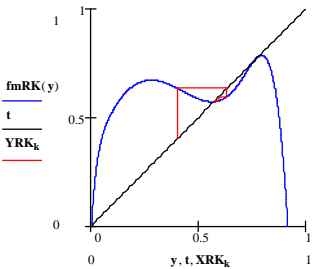

Time series

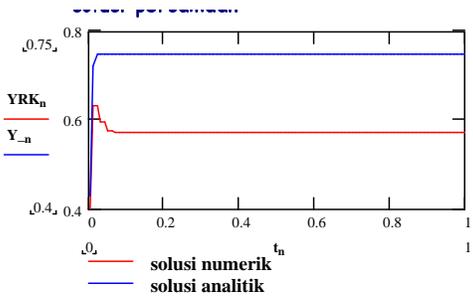

With that condition, new problem arises, i.e. if one differential equation can generate exact solution and some numerical method for the same problems produce chaos, then it is assumed that the exact solution is correct. What about differential equations that have no exact solution and the numerical method to solve them produce chaos? Can the observed system be called chaotic system? One of the example is Lorenz equation.

**Conclusion**

From above study, it can be explained that iteration method contribute to chaotic solution Some numerical method contain problems: iteration computation, error sampling frequency, rounding the result. To overcome those iteration, error sampling and rounding the result problems, one must provide non iterating method. It is known that neural network, fuzzy logic, cellular automata, etc are non-iterating method such as Euler, Predictor-Corrector, etc. Neural network, fuzzy logic and cellular automata are known as competent universal approximator as well as adaptive method.

Equations from universal approximation can be used to solve numerical problem of ODE and PDE. By minimizing the three numerical problems one can be sure that computing numerical process do not cause chaotic system.